\documentstyle[12pt]{article}
\def\be{\begin{equation}}
\def\ee{\end{equation}}
\def\bea{\begin{eqnarray}}
\def\eea{\end{eqnarray}}

\def\a{\alpha}
\def\b{\beta}

\begin{document}
\flushright{
\begin{center}
{\bf \Large{Solving Single and Many-body Quantum Problems: A Novel
Approach}}
\end{center}
\begin{center}
{$^*$\it N. Gurappa} \footnote{gurappa@ipno.in2p3.fr},
{\it Prasanta K. Panigrahi} \footnote{panisp@uohyd.ernet.in},\\
{\it R. Atre} \footnote{panisprs@uohyd.ernet.in} and
{\it T. Shreecharan}\\
$^*$Laboratoire de Physique Theorique et Modeles Statistiques, bat. 100,\\
Universit${\acute e}$ Paris-Sud-91405, Orsay, FRANCE\\
School of Physics, University of Hyderabad, Hyderabad,\\
Andhra Pradesh 500 046, INDIA
\end{center}
\abstract{A unified approach, for solving a wide class of single
and many-body quantum problems, commonly encountered in literature
is developed based on a recently proposed method for finding
solutions of linear differential equations. Apart from dealing
with exactly and quasi-exactly solvable problems, the present
approach makes transparent various properties of the familiar
orthogonal polynomials and also the construction of their
respective ladder operators. We illustrate the procedure for
finding the approximate eigenvalues and eigenfunctions of
non-exactly solvable problems.}

\section{Introduction}

The familiar series solution approach \cite{morse} is routinely
employed, when one encounters differential equations in quantum
mechanical eigenvalue problems. This approach, not only is tedious
to implement, but also throws little light on the underlying
symmetries of the problem at hand. Hence, one looks for alternate
simpler methods, a classic example being the elegant ladder
operator approach to the harmonic oscillator problem. Another
often used approach is the factorization technique pioneered by
Schr$\ddot o$dinger \cite{sch}, Infeld and Hull \cite{hull}.

In a number of cases, the underlying symmetries of the equation
under consideration, have led to algebraic approaches based on
group theory \cite{gursey} and supersymmetry \cite{sqm1}. Most of
these methods fail to generalize to many-body quantal systems and
are also not easily applicable to non-exactly solvable problems.

In this paper, we give a pedagogical description of a recently
developed method for solving linear differential equations
\cite{charan1,guru1} and apply it to a number of single and
many-body quantum problems. The proposed method is simple  and
assumes no special symmetry of the differential equation (DE)
under consideration. The symmetries and the algebraic structure of
the solution space emerge in a natural manner \cite{charan2}.

The paper is organized as follows. In the following section, we
give a brief description of the proposed method of solving linear
DEs, both, for the single and the many variable cases. For the
purpose of illustration, we then consider the familiar harmonic
oscillator problem, as also the related Hermite DE. Novel
expressions for the solutions of the confluent hypergeometric and
hypergeometric equations are then provided for later use.

Section 3 is devoted to the construction of the ladder operators.
It is shown that the novel form of the solutions, in conjunction
with the Baker-Campbell-Hausdorff (BCH) formula, lead to a
straightforward construction of the ladder operators, much akin to
the harmonic oscillator case. We then proceed to the quasi-exactly
solvable problem in section 4, wherein, the utility of the present
approach for finding approximate eigenvalues and eigenfunctions
for non-exactly solvable problems is also illustrated. Section 5
is devoted to the many-body correlated systems, which are
currently under intense study. The procedure and the subtleties
involved in dealing with such multi-variate Hamiltonians are
explicitly pointed out. We then conclude in section 6 after
pointing out other areas, where this approach may find fruitful
application.

 For the purpose of facilitating
comparison with the standard literature, units have been
appropriately chosen in different sections.

\section{A simple approach to familiar differential equations}

For simplicity, we will first consider the case of single variable
linear DEs and point out its multi-variate generalization later. A
single variable linear DE, as will become clear from the examples
of later sections, can be cast in the form
\begin{equation} \label{ie}
\left[F(D) + P(x,d/dx)\right] y(x) = 0 \quad,
\end{equation}
where, $D \equiv x d/dx$ is the Euler operator, $F(D) \equiv
\sum_{n = - \infty}^{n = \infty} a_n D^n $ and $a_n$'s are some
parameters; $P(x,d/dx)$ can be an arbitrary polynomial function of
$x,d/dx$ and other operators. The solution to Eq. (\ref{ie}) can
be written as \cite{charan1,guru1},
\begin{equation} \label{an}
y(x) = C_\lambda \left \{\sum_{m = 0}^{\infty} (-1)^m
\left[\frac{1}{F(D)} P(x,d/dx)\right]^m \right \} x^\lambda
\end{equation}
provided, $F(D) x^\lambda = 0$; here $C_\lambda$ is constant.
Before proceeding further, we list two important properties of the
Euler operator to be extensively used in the text. Euler operator
is diagonal in the space of monomials i.e., $Dx^\lambda = \lambda
x^\lambda$ and other operators carry definite degrees, with
respect to the Euler operator i.e., $[D,O^d] = d\,O^d$, where $d$
is the degree of the operator $O^d$. For example, $[D,x^2]=2\,x^2$
and $[D,d^2/dx^2] = -2\,d^2/dx^2$. Using the above results, it is
easy to see that, the operator $1/F(D)$ is well defined in the
above expression and will not lead to any singularity, if
$P(x,d/dx)$ does not contain any degree zero operator.

The proof of Eq. (\ref{an}) is straightforward and follows by
direct substitution \cite{charan1}. Alternatively, since
$F(D)x^{\lambda}=0$, equating $F(D)x^{\lambda}$ modulo
$C_{\lambda}$ and Eq. (\ref{ie}), one finds,
\begin{equation}
\left[ F(D) + P(x,d/dx)\right] y(x) = C_{\lambda} F(D) x^{\lambda} \quad.
\end{equation}
Rearranging the above equation in the form
\begin{equation}
F(D) \left[ 1 + \frac{1}{F(D)} P(x, d/dx) \right] y(x) =
C_{\lambda} F(D) x^{\lambda} \quad,
\end{equation}
and cancelling $F(D)$, we obtain
\begin{equation}
\left[1 + \frac{1}{F(D)} P(x, d/dx) \right]y(x) =
C_{\lambda}x^{\lambda}\quad.
\end{equation}
This yields
\begin{equation}
y(x) = C_{\lambda} \frac{1}{\left[1+
\frac{1}{F(D)}P(x,d/dx)\right]} x^{\lambda}\quad,
\end{equation}
which can be cast in the desired series form: \bea \nonumber y(x)
= C_{\lambda} \sum^{\infty}_{m=0}(-1)^m
\left[\frac{1}{F(D)}P(x,d/dx)\right]^m x^{\lambda}\quad. \eea It
is explicit that, the above procedure connects the solution $y(x)$
to the space of the monomials $x^\lambda$. This fact will be
exploited in the later sections for obtaining the ladder operators
and explicate various properties of the solution space.

The generalization of this method to a wide class of many-variable
problems is immediate. Using the fact that, $F(\bar D) X^\lambda =
0$ has solutions, in the space of monomial symmetric functions
\cite{sf}, where $\bar D = \sum_iD_i \equiv \sum_i x_i
\frac{d}{dx_i}$, the solutions of those multi-variate DEs, which
can be separated into the form given in Eq. (\ref{ie}), can be
solved like the single variable case. As will be seen later, this
procedure enables one to solve a number of correlated many-body
problems.

For illustration, we consider the harmonic oscillator problem. The
Schr$\ddot{o}$dinger eigenvalue equation (in the units,
$\hbar$=$\omega$=$m$=1)
\begin{equation}
\left[\frac{d^2}{d x^2} + (2 E_n - x^2)\right] \psi_n = 0 \quad,
\end{equation}
can be written in the form given in Eq. (\ref{ie}), after
multiplying it by $x^2$ :
\begin{equation}
\left[(D - 1) D + x^2 (2 E_n - x^2)\right] \psi_n = 0 \quad.
\end{equation}
Here, $F(D)=(D - 1)D$ and the condition $F(D) x^{\lambda} = 0$
yields, $\lambda = 0$ or $1$. Using Eq. (\ref{an}), the solution
for $\lambda = 0$ is,
\begin{eqnarray}
\psi_0 &=& C_0 \left \{\sum_{m = 0}^{\infty} (-1)^m
\left[\frac{1}{(D - 1) D} (x^2 (2 E_0 - x^2))\right]^m \right \}
x^0 \nonumber\\ \nonumber\\
&=& C_0 \left[1 - \frac{[2 E_0]}{2 !} x^2 + \frac{(2! + [2
E_0]^2)}{4!} x^4 - \frac{(4! + (2!)^2 [2 E_0] + (2!) [2
E_0]^3)}{2! 6!} x^6 + \cdots \right] \,\,.
\end{eqnarray}
Here $\psi_0$ is an expansion in powers of $x$, whose coefficients
are polynomials in $E_0$. The above series can be written in a
closed, square integrable form, $C_0 \exp(- x^2/2)$, only when
$E_0=1/2$. Analogously, $\lambda=1$, yields the first excited
state. To find the $n^{\mathrm th}$ excited state, one has to
differentiate the Schr${\ddot o}$dinger equation $(n-2)$ number of
times and subsequently multiply it by $x^n$ to produce a $F(D)=
x^n \frac{d^n}{d x^n} = \prod_{l = 0}^{n - 1} (D - l)$, and
proceed in a manner similar to the ground state case.

It is clear that, our procedure yields a series solution, where
additional conditions like square integrability has to be imposed
to obtain physical eigenfunctions and their corresponding
eigenvalues. Once, the ground state has been identified and for
those cases, where $\psi(x)= \psi_0P(x)$, where $P(x)$ is a
polynomial, one can effortlessly obtain the polynomial part, as
will be shown below. Proceeding with the harmonic oscillator case
and writing \bea\nonumber\psi_\a(x) =
\exp(-\frac{x^2}{2})H_{\alpha}(x)\quad,\eea one can easily show
that $H_\a$ satisfies\be \label{h}\left[D - \alpha - \frac{1}{2}
\frac{d^2}{dx^2} \right] H_{\alpha}(x) = 0 \quad, \ee where
$\alpha = E_n-1/2$. The solution of the DE \be \label{shree}
H_{\alpha}(x) = C_{\alpha} \sum_{m = 0}^{\infty}(-1)^m
\left[-\frac{1}{(D - \alpha)}\frac{1}{2} \frac{d^2}{dx^2} \right
]^{m} x^\a \quad.\ee yields a polynomial only when $\a$ is an
integer, since the operator $d^2/dx^2$ reduces the degree of
$x^\a$ by two, in each step. Setting $\a=n$ in Eq. (\ref{h}), we
obtain the Hermite DE and $E_n =(n+1/2)$ as the energy eigenvalue.
Below, we give the algebraic manipulations required to cast the
series solution of Eq. (\ref{shree}) into a form, not very
familiar in the literature. For the Hermite DE, $F(D)=D - n$ and
$P(x,d/dx)=-\frac{1}{2}\frac{d^2}{dx^2}$, the condition $F(D)
x^\lambda = 0$ yields $\lambda = n$, hence,
\begin{eqnarray} \label{sol} H_n(x) = C_n
\sum_{m = 0}^{\infty}(-1)^m \left[-\frac{1}{(D - n)} \frac{1}{2}
\frac{ d^2}{dx^2} \right ]^{m} x^n \quad.
\end{eqnarray}
Using, $[D \,,(d^2/dx^2)] = -2 (d^2/dx^2)$ and making use of the
fact \be \frac{1}{(D-n)} = {\int_0}^\infty ds \, e^{-s(D-n)} \ee
we can write,
\begin{eqnarray} \nonumber
\left[- \frac{1}{2} \frac{1}{(D-n)} \frac{d^2}{dx^2} \right]&=&
- \frac{1}{2} \frac{d^2}{dx^2} \frac{1}{(D-n-2)} \quad, \nonumber \\
\nonumber \\
\left[- \frac{1}{2} \frac{1}{(D-n)} \frac{d^2}{dx^2} \right]
\left[- \frac{1}{2} \frac{1}{(D-n)} \frac{d^2}{dx^2} \right]&=&
\left[ - \frac{1}{2} \frac{d^2}{dx^2}\right]^2 \frac{1}{(D-n-4)}
\frac{1}{(D-n-2)} \quad.
\end{eqnarray}
Hence in general,
\begin{eqnarray} \label{rel}
\left[- \frac{1}{2}\frac{1}{(D - n)}\frac{d^2}{dx^2} \right ]^{m}
x^n &=& \left(- \frac{1}{2}\frac{d^2}{dx^2}\right)^m
\prod_{l=1}^m \frac{1}{(- 2 l)} x^n \quad, \nonumber \\
&=& \frac{1}{m!} \left(\frac{1}{4} \frac{d^2}{dx^2}\right)^m x^n
\quad.
\end{eqnarray}
Substituting Eq. (\ref{rel}) in Eq. (\ref{sol}), we obtain,
\begin{eqnarray}
H_n(x) &=& C_n \sum_{m = 0}^{\infty}(-1)^m \frac{1}{m!}
\left(\frac{1}{4}\frac{d^2}{dx^2}\right)^m
x^n  \quad, \nonumber\\ \label{s}  \nonumber\\
&=& C_n \exp{\left(- \frac{1}{4} \frac{d^2}{dx^2}\right)} x^n \quad,
\end{eqnarray}
a result, not commonly found in the literature \cite{fer}. The
arbitrary constant $C_n$ is chosen to be $2^n$, so that the
polynomials obtained can match with the standard definition
\cite{grad}.

The algebraic manipulations shown above can be applied to the
confluent hypergeometric DE
\begin{equation}\label{chg}
\left[x \frac{d^2}{dx^2}+ (\gamma - x) \frac{d}{dx} - \alpha
\right] \Phi(\alpha; \gamma; x) = 0 \quad,
\end{equation}
to give
\begin{equation}
\Phi(\alpha,\gamma,x)= (-1)^{ -\alpha}
\frac{\Gamma(\gamma)}{\Gamma(\gamma-\alpha)} \exp{\left( -x
\frac{d^2}{dx^2}-\gamma \frac{d}{dx}\right)}\,.\, x^{- \alpha}
\quad,
\end{equation}
where the normalization has been chosen appropriately. Likewise,
the solution to the hypergeometric DE
\begin{equation}
\left[x^2 \frac{d^2}{dx^2} + {\left(\alpha + \beta + 1 \right) x
\frac{d}{dx}} + \alpha\beta - x \frac{d^2}{dx^2} - \gamma
\frac{d}{dx} \right] F{(\alpha, \beta; \gamma; x)}= 0
\end{equation}
can be written as,
\begin{equation}
F(\alpha, \beta; \gamma; x)= (-1)^{-\beta} { \frac{\Gamma ( \alpha
- \beta)\Gamma(\gamma)} { \Gamma (\gamma - \beta) \Gamma(\alpha)}
\exp{\left[\frac{-1}{\left(D+\alpha \right)} \left(x
\frac{d^2}{dx^2}+ \gamma \frac{d}{dx}\right)
\right]}}\,.\,x^{-\beta} \quad.
\end{equation}
For convenience a table has been provided at the end, which lists
a number of commonly encountered DEs and the novel exponential
forms of their solutions. It is worth pointing out that, the
solutions of confluent hypergeometric and hypergeometric DEs given
above, are polynomial solutions, provided $\a$ and $\b$ are
negative integers. It should be noticed that, unlike the
conventional expressions, the monomials in the above solutions are
arranged in decreasing powers of $x$.

The series solutions for the same can be obtained by a simple
modification of the DE. Multiplying Eq. (\ref{chg}) by $x$, we get
\be\left[x^2 \frac{d^2}{dx^2}+ \gamma x \frac{d}{dx}-
x^2\frac{d}{dx} - \alpha x \right] \Phi(\alpha; \gamma; x) = 0
\quad,\ee for which, $F(D)= \left(x^2 \frac{d^2}{dx^2}+ \gamma x
\frac{d}{dx}\right)$ and $P(x,d/dx) = - x^2\frac{d}{dx} - \alpha
x$. The requirement $F(D)x^\lambda = 0$, generates two roots,
$\lambda = 0, 1-\gamma$, yielding the two linearly independent
solutions. For $\lambda = 0$, we obtain the familiar confluent
hypergeometric series \be \Phi(\alpha; \gamma; x)= 1 +
\frac{\alpha}{\gamma}x+ \frac{\alpha(\alpha+1)}{\gamma(\gamma+1)}
\frac{x^2}{2!}+ \cdots \quad.\ee Similarly multiplying the
hypergeometric DE with $x$ yields two roots, $\lambda = 0,
1-\gamma$ and $\lambda = 0$ solution gives rise to the well-known,
Gauss hypergeometric series. Since a number of quantum mechanical
problems can be related to confluent and hypergeometric DEs
\cite{landau}, we hope that the novel expressions given above and
 in the table will find physical applications.

\section{Construction of Ladder Operators}

In this section, we first derive the ladder operators for the
harmonic oscillator, making use of the exponential form of the
solution for the Hermite polynomials derived earlier and then
proceed to the Coulomb problem and the Laguerre polynomials
associated with it. Subsequently, we outline the steps required
for generalizing these results for other cases.

The advantage of the present approach, which connects the solution
space to the space of the monomials, lies in the fact that, at the
level of the monomials, the ladder operators can be constructed
easily, which facilitates the construction of the same at the
level of the polynomials and the wave functions. The raising and
lowering operators for the Hermite polynomials are derived,
starting respectively from $x$ and $d/dx$ at the level of the
monomials. For example, $xx^n=x^{n+1}$, when operated by
$\exp[-(1/4)(d^2/dx^2)] \equiv \exp(-A)$, from the left on both
the sides and after introducing an identity operator suitably,
yields
\begin{equation}
e^{-A} 2x e^{A}2^n e^{-A}x^n = 2^{n+1} e^{-A}x^{n+1} \quad,
\end{equation}
which leads to
\begin{equation}
\left[2x-\frac{d}{dx}\right]H_n(x)=H_{n+1}(x) \quad.
\end{equation}
In deriving the above relation, we have made use of the BCH
formula,
\begin{equation}
e^{-A}Be^{A}=
B + [B,A] + \frac{1}{2!} \left[[B,A],A \right] + \cdots \quad.
\end{equation}

The construction of the lowering operator is analogous to the
above procedure. From $\frac{d}{dx} x^n = n x^{n-1}$, one obtains,
by performing a similarity transformation,
\begin{equation}
e^{-A} \frac{d}{dx} e^{A}2^n e^{-A} x^n
= 2n2^{n-1}e^{-A} x^{n-1} \quad,
\end{equation}
or
\begin{equation}
\frac{d}{dx}H_n(x)=2n H_{n-1}(x) \quad.
\end{equation}
The creation and annihilation operators for the harmonic
oscillator problem follow after one more similarity
transformation:
\begin{equation}
e^{-\frac{x^2}{2}}
[2x-\frac{d}{dx}]e^{\frac{x^2}{2}}\quad
\sqrt{\frac{1}{2^nn!}}e^{-\frac{x^2}{2}}H_n(x) =
\sqrt{\frac{2(n+1)}{2^{n+1}(n+1)!}}e^{-\frac{x^2}{2}}H_{n+1}(x)\quad,
\end{equation}
or
\begin{equation}
\left[x-\frac{d}{dx}\right]\psi_n(x) = \sqrt{2(n+1)}
\psi_{n+1}(x) \quad;
\end{equation}
and
\begin{equation}
e^{-\frac{x^2}{2}}
\frac{d}{dx}e^{\frac{x^2}{2}}\quad
\sqrt{\frac{1}{2^nn!}}e^{-\frac{x^2}{2}}H_n(x) =
\sqrt{\frac{2n}{2^{n-1}(n-1)!}}e^{-\frac{x^2}{2}}H_{n-1}(x)\quad,
\end{equation}
leads to
\begin{equation}
\left[\frac{d}{dx} + x \right] \psi_n(x)
 = \sqrt{2n} \psi_{n-1}(x) \quad.
\end{equation}
Here\be
\psi_n=\sqrt{\frac{1}{2^n}}\exp(-\frac{x^2}{2})H_n(x)\quad.\ee

Proceeding in an analogous manner, we derive the ladder operators
for the Coulomb problem. The polynomial solution of the Laguerre
DE
\begin{equation}
\left[ x \frac{d^2}{dx^2}+ (\alpha - x + 1)\frac{d}{dx}+ n
\right]L^{\alpha}_n(x)=0 \quad,
\end{equation}
can be written as, \be L^{\alpha}_n(x)= \frac{(-1)^n}{n!} \exp
\left[ - x \frac{d^2}{dx^2} - (\alpha+1) \frac{d}{dx}
\right]\,.\,x^n \quad.\ee With $B=[x {d^2}/{dx^2}+(\alpha+1)d/dx]$
and proceeding as in the Hermite polynomial case we obtain
\begin{equation}
\left[x-2x \frac{d}{dx}-(\alpha+1)+ x \frac{d^2}{dx^2}
+(\alpha+1) \frac{d}{dx}\right]L^{\alpha}_n(x)
 = -(n+1)L^{\alpha}_{n+1}(x) \quad.
\end{equation}
To construct the lowering operator, one should start with a
lowering operator different from $d/dx$ at the level of the
monomials, since a similarity transformation on this operator does
not lead to a closed form expression. The simplest one is the
operator $B$ itself, since it commutes with $e^{-B}$. Acting $B$
on the monomial $x^n$, we obtain
\begin{equation}
\left[x \frac{d^2}{dx^2}+(\alpha+1) \frac{d}{dx}\right]
x^n = n(n+\alpha)x^{n-1} \quad,
\end{equation}
and performing a similarity transformation through $e^{-B}$, one
gets
\begin{equation}
\left[x \frac{d^2}{dx^2}+(\alpha+1) \frac{d}{dx}\right]
L^{\alpha}_n(x) = -(n+ \alpha)L^{\alpha}_{n-1}(x)\quad.
\end{equation}
To obtain the ladder operators for the radial wave functions, $R
^{\alpha} _n(x)=x^le^{-\frac{x}{2}}L^{\alpha}_n(x)$ of the Coulomb
Hamiltonian, \cite{landau} one proceeds in a manner similar to the
oscillator problem. Explicitly, the action of the raising and
lowering operator, respectively are,
\begin{eqnarray}
\left[\frac{x}{2}-x \frac{d}{dx} + x \frac{d^2}{dx^2} + 2
\frac{d}{dx} -\frac{l(l+1)}{x} - 1 \right]R^{\alpha}_n(x)=
~-(n+1)R^{\alpha}_{n+1}(x) \quad,
\end{eqnarray}
and
\begin{eqnarray}
\left[ \frac{x}{2}+ x \frac{d}{dx}+ x \frac{d^2}{dx^2}+
2\frac{d}{dx} - \frac{l(l+1)}{x}+1\right]
R^{\alpha}_n(x)=~-(n+\alpha)R^{\alpha}_{n-1}(x) \quad.
\end{eqnarray}
It is worth pointing out that the value of $\a=(2l+1)$ can also be
changed, since
\begin{equation}
\frac{d}{dx}L^{\alpha}_n(x) = (-1^n/n!)\exp
\left[-x\frac{d^2}{dx^2} -
(\alpha+2)\frac{d}{dx}\right]\frac{d}{dx}x^n \quad,
\end{equation}
from where, one obtains the standard result,
\begin{equation}
\frac{d}{dx}L^{\alpha}_n(x) = - L^{\alpha+1}_{n-1}(x) \quad.
\end{equation}
It is clear that a composite operator can be obtained keeping $n$
unchanged, while changing the value of $\alpha$. The forms of the
ladder operators given above are not unique, one can construct
more complicated operators starting from different ones at the
monomial level.

The exponential form of the solutions enables one to construct
ladder operator for hypergeometric, confluent hypergeometric and
other polynomials and functions \cite{charan2}. which can be used
to construct similar operators for the quantum mechanical problems
associated with these polynomials, e.g., Morse, P$\ddot
o$schl-Teller, Eckart and other potentials \cite{landau,text}.

\section{Quasi-Exactly Solvable Problems}

This section is devoted to the study of QES problems
\cite{shifman,ush}. These problems are intermediate to exactly and
non-exactly solvable quantum potentials, in the sense that, only a
part of the spectrum can be determined analytically. These
potentials have attracted considerable attention in recent times,
because of their connection to various physical problems.

We illustrate our procedure, through the sextic oscillator (in the
units $\hbar=2m=\omega=1$), whose eigenvalue equation is given by
:
\begin{equation}
\left[- \frac{d^2}{dx^2}+ \alpha x^2 +\gamma x^6 \right])\psi(x) =
E \psi(x) \quad.
\end{equation}
 Asymptotic analysis suggests a trial wave function of the form,
\be \psi(x)=\exp(-bx^4){\tilde \psi}(x) \quad,\ee which leads to,
\be\label{sex2}
\left[-\frac{d^2}{dx^2}+2{\sqrt\gamma}x^3\frac{d}{dx}+(\alpha+3{\sqrt\gamma})x^2
 \right]{\tilde \psi}(x)= E\tilde{\psi}(x) \quad,\ee where
 $x^6$ term has been removed by the condition $16b^2=\gamma$.
One notices that, the operator ${\tilde O} =
(\alpha+3{\sqrt\gamma})x^2 +2{\sqrt\gamma}x^3 d/dx$ increases the
degree of ${\tilde \psi}(x)$ by two, if ${\tilde \psi}(x)$ is a
polynomial. Confining ourselves to polynomial solutions and
assuming that the highest power of the monomial in ${\tilde
\psi}(x)$ is $n$, one obtains, \be
-\frac{\alpha}{\sqrt\gamma}=2n+3\ \quad ,\ee  after imposing the
condition that $\tilde{O}$ does not increase the degree of the
polynomial. This is the well-known relationship between the
coupling parameters of the quasi-exactly solvable sextic
oscillator. Taking $n=4$ and $\gamma=1$ for simplicity, and after
multiplying the above equation with $x^2$ : \be \label{sex2}\left[
D(D-1) + Ex^2 + 8x^4 - 2x^5\frac{d}{dx}\right]{\tilde \psi}(x)= 0
 \quad ,\ee we get,
 \be {\tilde \psi}_{0}(x)=C_0
\left\{ \sum_{m=0}^{\infty}(-1)^m
\left[\frac{1}{D(D-1)}\left(x^2E_0+8x^4-2x^5\frac{d}{dx}\right)\right]^m
\right\}\,1 \quad.\ee Modulo $C_0$, the above series can be
expanded as,\be{\tilde \psi}_{0}(x) = 1 - E_0\frac{x^2}{2!} +
(E_0^2 - 16)\frac{x^4}{4!}+(64E_0 - E_0^3)\frac{x^6}{6!} + \cdots
\quad.\ee The monomials having degree greater than four vanish
provided, $E_0 = 0,\pm 8$. It can be explicitly checked that, for
these values of $E_0$, Eq. (\ref{sex2}) is satisfied. The
eigenfunctions corresponding to these three values are given by,
\bea \psi_{-8}(x) &=&
\exp(-\frac{x^4}{4})[1+ 4x^2 + 2x^4] \\
\psi_0(x) &=& \exp(-\frac{x^4}{4})[1- \frac{2}{3}x^4] \quad,\\
{\mathrm and} \qquad\psi_{+8}(x) &=& \exp(-\frac{x^4}{4})[1- 4x^2
+ 2x^4]\quad.\eea This procedure generalizes to a wide class of
QES problems \cite{atre}.

Below, we demonstrate the method of finding approximate
eigenvalues and eigenfunctions for non-exactly solvable problems,
using the well studied anharmonic oscillator as the example :
\be\label{anhar1} \left[-\frac{d^2}{dx^2}+\alpha x^2 + \beta x^4 -
E_n\right]\psi_n(x)=0 \quad.\ee Proceeding as before, $\psi_0(x)$
can be written as,
\begin{eqnarray}
\psi_0(x) = C_0 \left \{\sum_{m = 0}^{\infty} (-1)^m
\left[\frac{1}{(D - 1) D}  {\left(E_0 x^2 -\alpha x^4 - \beta
x^6\right)}\right]^m \right \}\,.\, 1 \quad ,
\end{eqnarray} which can be expanded as,
\begin{eqnarray}\label{anhar}
\psi_{0}(x)=1-\frac{
E_0}{2!}x^2+\frac{1}{4!}\left(2\alpha+E_0^2\right)x^4-
\frac{1}{6!}\left(24 \beta-(14\alpha E_0+E_0^3)\right)x^6+
\cdots\end{eqnarray} Although a number of schemes can be devised
for the purpose of approximation, we consider the simplest one of
starting with a trial function ${\tilde\psi}_0(x)= \exp(-\mu
x^2-\nu x^4 )$ and matching it with $\psi_0(x)$. Comparison of the
first three terms yields,
\bea \nonumber \mu&=& \frac{E_0}{2!} \quad,\\
\nonumber \frac{\mu^2}{2!} -
\nu&=&\frac{2\alpha}{4!}+\frac{E_0^2}{4!}\quad,\\
{\mathrm and}\qquad \mu\nu- \frac{\mu^3}{3!}&=& \frac{\beta}{30} -
\frac{(14E_0+E_0^3)}{6!}\quad.\eea The resulting cubic equation in
energy, $E_{0}^{3}- E_0\alpha={3\beta}/2$, leads to one real root
and two complex roots. Choosing the real root on physical grounds,
one obtains,\be E_0=\frac{2^{1/3}\alpha}{A}+\frac{A}{{3}\cdot
2^{1/3}}\quad,\ee where
$A=\left[40.5\beta+(1640.25\beta^2-108\alpha^3)^{1/2}\right]^{1/3}$.
The value of $E_0$, obtained in the weak coupling regime, matches
reasonably well with the earlier obtained results \cite{mfar}. An
approximate $\psi_0$ can be obtained from Eq. (\ref{anhar}). One
can easily improve upon the above scheme by taking better trial
wave functions. Similar analysis can be carried out for the
excited states. The above expansion of the wave function may be
better amenable for a numerical treatment. For example, an
accurate numerically determined energy value can lead to a good
approximate wave function.

\section{Many-body interacting systems}

In this section, we will be dealing with correlated many-body
systems, particularly of the Calogero-Sutherland \cite{calo} and
Sutherland type \cite{suth}. These models have found application
in diverse branches of physics like fluid flow, random matrix
theory, novel statistics, quantum Hall effect and others
\cite{simon,g}. A number of methods e.g., Lax pair \cite{lax},
Bethe-ansatz techniques \cite{bethe} and $S_{N}$-extended
Heisenberg algebra \cite{guru1,vasi,polyc} have been employed for
studying these systems.

We start with the relatively difficult Sutherland model, where the
particles are confined to a circle of circumference $L$. The
two-body problem treated explicitly below, straight-forwardly
generalizes to $N$ particles. The Schr\"odinger equation is given
by (in the units $\hbar=m=1$)

\begin{eqnarray} \label{sut1}
\left[- \frac{1}{2}\sum_{i=1}^2 \frac{\partial^2}{ \partial x_i^2}
+ \beta (\beta - 1) \frac{\pi^2}{L^2} \frac{1}{\sin^2[\pi (x_1 -
x_2)/L]} - E_\lambda \right] \psi_\lambda(\{x_i\}) = 0 \qquad.
\end{eqnarray}
Taking, $z_j = e^{2\pi i x_j /L}$ and writing
$\psi_\lambda(\{z_i\}) = \prod_{i,{i \neq j}} z_i^{-\beta
/2}(z_i-z_j)^\beta J_\lambda(\{z_i\})$, the above equation
becomes,
\begin{eqnarray} \label{jac}
\left[\sum_{i=1}^2 D_i^2 + \beta \frac{z_1 + z_2}{z_1 - z_2} (D_1
- D_2) + \tilde{E}_0 - \tilde{E}_\lambda \right]
J_\lambda(\{z_i\}) = 0 \qquad,
\end{eqnarray}
where, $D_i \equiv z_i \frac{\partial}{\partial z_i}$,
$\tilde{E_\lambda} \equiv 2(\frac{L}{2 \pi})^2 E_\lambda$,
$\tilde{E_0} \equiv 2(\frac{L}{2 \pi})^2 E_0$ and $E_0 =
(\frac{\pi}{L})^2 \beta^2$, is the ground-state energy. Here,
$J_\lambda(\{z_i\})$ is the polynomial part, which in the
multivariate case is the well known Jack polynomial \cite{sf}.
Here, $\lambda$ is the degree of the symmetric function and
$\{\lambda\}$ refers to different partitions of $\lambda$. $\sum_i
D_i^2$ is a diagonal operator in the space spanned by the monomial
symmetric functions, $m_{\{\lambda\}}$, with eigenvalues
$\sum_{i=1}^2 \lambda_i^2$ . A monomial symmetric function is a
symmetrized combination of monomials of definite degree. For
example for two particle case, there are two monomial symmetric
functions having degree two. These are $m_{2,0}=
x_{1}^{2}+x_{2}^{2}$ and $m_{1,1}=x_{1}x_{2}$. Readers are
referred to Ref. [9] for more details about various symmetric
functions and their properties.
 Rewriting Eq. (\ref{jac}) in the form,
\begin{eqnarray}
\left[\sum_i (D_i^2 - \lambda_i^2) + \beta \frac{z_1 + z_2}{z_1 -
z_2} (D_1 - D_2) + \tilde{E}_0 + \sum_i \lambda_i^2 -
\tilde{E}_\lambda \right] J_\lambda(\{z_i\}) = 0 \qquad,
\end{eqnarray}
one can immediately show that,
\begin{eqnarray} \label{sr}
J_\lambda(\{z_i \}) &=& C_\lambda \left\{\sum_{n = 0}^{\infty}
(-1)^n \left[\frac{1}{\sum_i (D_i^2 - \lambda_i^2)}\left(\beta
\frac{z_1 + z_2}{z_1 - z_2}(D_1 - D_2) +  \tilde{E}_0 + \sum_i
\lambda_i^2 - \tilde{E}_\lambda\right)\right]^n \right\} \nonumber\\
&& \qquad \qquad \qquad \qquad \qquad \qquad \qquad \qquad \qquad
\qquad \qquad \qquad  \times  m_\lambda(\{z_i\})\quad .
\end{eqnarray}
For the sake of convenience, we define
\begin{eqnarray} \label{S}
\hat{S} & \equiv & \left[\frac{1}{\sum_i (D_i^2 - \lambda_i^2)}
\hat{Z} \right] \qquad, \nonumber\\
\mbox{and} \qquad \hat{Z} & \equiv & \beta \frac{z_1 + z_2}{z_1 -
z_2} (D_1 - D_2) +  \tilde{E}_0 + \sum_i \lambda_i^2 -
\tilde{E}_\lambda \qquad.
\end{eqnarray}
The action of $\hat{S}$ on $m_\lambda(\{z_i \})$ yields
singularities, unless one chooses the coefficient of $m_\lambda$
in $\hat{Z} \, m_\lambda(\{z_i\})$ to be zero; this condition
yields the eigenvalue equation
$$
\tilde{E}_\lambda = \tilde{E}_0 + \sum_i (\lambda_i^2 + \beta [3-
2 i] \lambda_i) \qquad.
$$
Using the above, one can write down the two particle Jack
polynomial as,
\begin{eqnarray} \label{nfj}
J_\lambda(\{z_i \}) = \sum_{n=0}^\infty (- \beta)^n
\left[\frac{1}{\sum_i (D_i^2 - \lambda_i^2)}(\frac{z_1 + z_2}{z_1
- z_2}(D_1 - D_2) - \sum_i
(3 - 2 i) \lambda_i )\right]^n \nonumber\\
\qquad \qquad \qquad \qquad \qquad  \times m_\lambda(\{x_i\})
\quad.
\end{eqnarray}
Starting from $m_{2,0} = z_1^2 + z_2^2$, it is straightforward to
check that
\begin{eqnarray}
\hat{Z} m_{2,0} &=& 4\beta  (z_1+ z_2)^{2} = 4 \beta m_{1,0}^{2} 
\nonumber\\
\hat{S} m_{2,0} &=& \frac{1}{\sum_i(D_i^2 - 4)}(4 \beta
m_{1,0}^{2})=-2
\beta m_{1,0}^{2} \qquad, \nonumber\\
\hat{S}^n  m_{2,0} &=& - 2 (\beta)^n m_{1,0}^{2} \quad
\mbox{for}\quad n \ge 1 \quad. \nonumber
\end{eqnarray}
Substituting the above result in Eq. (\ref{sr}), apart from $C_2$,
one obtains
\begin{eqnarray}
J_{2} &=& m_{2,0} + \left(\sum_{n=1}^\infty (- 1)^n (-2) (\beta)^n
\right)
m_{1,0}^{2} \nonumber\\
&=& m_{2,0} + 2 \beta \left(\sum_{n = 0}^\infty (- \beta)^n
\right)
m_{1,0}^{2} \nonumber\\
&=& m_{2,0} + \frac{2 \beta}{1 + \beta} m_{1,0}^{2} \qquad;
\end{eqnarray}
which is the desired result. The above approach can be easily
generalized to the $N$-particle case.

Another class of many-body problems, which can be solved by the
present approach is the Calogero-Sutherland model (CSM) and its
generalizations \cite{lax}. Proceeding along the line, as for the
Sutherland model, one finds the eigenvalues and eigenfunctions for
the CSM.

The Schr\"odinger equation for the CSM in the previous units, is
given by, \bea
{\left[{-\frac{1}{2}\sum_{i=1}^{N}\frac{\partial}{\partial
{x_{i}}^2} + \frac{1}{2}\sum_{i=1}^{N}x_{i}^{2}+
\frac{g^2}{2}\sum_{{i,j}\atop {i\ne
j}}^{N}{\frac{1}{(x_{i}-x_{j})^2}}}
-E_{n}\right]}\psi_{n}(\{x_i\})=0 \quad,\eea where the wave
function is of the form, \cite{calo} \bea
\psi_{n}(x)=\psi_0P_n(\{x_i\})=ZGP_n(\{x_i\})\quad.\eea Here
$Z\equiv\prod_{i<j}(x_{i}-x_{j})^{\beta}$, $G \equiv
\exp\left\{-\frac{1}{2}\sum_{i}x_{i}^{2}\ \right\}\,$, $g^2 =
\beta(\beta-1)$ and $P_n(\{x_i\})$ is a polynomial. After removing
the ground state, the polynomial $P_n(\{x_i\})$ satisfies,
\begin{eqnarray} \label{til}
\left[\sum_i x_i \frac{\partial} {\partial x_i} +  E_0 - E_n -
\frac{1}{2}\sum_{i}\frac{\partial^2}{\partial{x_{i}}^{2}}-\beta\sum_{i\neq
j}\frac{1}{(x_{i}-x_{j})}\frac{\partial}{\partial{x_{i}}}\right]
P_n(\{x_i\}) = 0 \quad,
\end{eqnarray} where
$E_{0}=\frac{1}{2}N+\frac{1}{2}{\beta}N(N-1)$. Defining \bea
\hat{A}(\beta) \equiv
\frac{1}{2}\sum_{i}\frac{\partial^2}{\partial{x_{i}}^{2}}+\beta\sum_{i\neq
j}\frac{1}{(x_{i}-x_{j})}\frac{\partial}{\partial{x_{i}}} \quad,
\nonumber\eea one can easily see, following the procedure adopted
for the Hermite DE that \be P_n(\{x_i\}) = C_n
e^{-\hat{A}(\beta)}m_{\{n\}}(\{x_i\}) \quad,\ee where
$m_{\{n\}}(\{x_i\})$ is a monomial symmetric function of degree
$n$. The corresponding energy is given by \bea E_{n}= E_{0}+n
\quad , \eea One can use the above procedure to solve many other
interacting systems.

\section{Conclusions}

In conclusion, we have presented a novel scheme to treat exactly,
quasi-exactly and non-exactly solvable problems, which also
extends to a wide class of many-body interacting systems. The
procedure was used for the construction of the ladder operators
for various orthogonal polynomials and the quantum systems
associated with them. The approximation scheme presented needs
further refinement. It should be analyzed in conjunction with
computational tools for finding its efficacy as compared to other
methods. The many-body problems presented here have deep
connection with diverse branches of physics and mathematics. The
fact that the procedure employed for solving them, connects the
solution space of the problem under study to the space of
monomials, will make it useful for constructing ladder operators
for the many-variable case. This will throw light on the structure
of the Hilbert space of these correlated systems. Some of these
questions are currently under study and will be reported
elsewhere. \vskip0.5cm {\bf Acknowledgements:} T. S. thanks U.G.C
(India) for financial support through the JRF scheme.

}
\newpage
\noindent{\small {\bf TABLE I. Some frequently encountered DE and
their novel solutions.}\\
Below, the differential equations from top to bottom,
respectively, are Hermite, Laguerre, Legendre, Gegenbauer,
Chebyshev Type I, Chebysgev Type II, Bessel, Confluent
Hypergeometric and Hypergeometric.
\begin{center}
\begin{tabular}{|c|c|c|}
\hline \hline
{\bf Differential Equation} & {\bf F(D)}, $D\equiv x(d/dx)$ & {\bf 
Solution}\\
\hline &&\\
 $\left[x\frac{d}{dx} - n
-\frac{1}{2}\frac{d^2}{dx^2}\right]H_n(x)=0$ & $(D - n)$ &
$H_n(x)= C_n \exp{\left[-\frac{1}{4}\frac{d^2}{dx^2}
\right]}. x^n$\\
&&\\
 $\left[x\frac{d}{dx}-n -(\alpha+1)\frac{d}{dx}-x\frac{d^2}{dx^2}
    \right]L^{\alpha}_n=0$
    & $(D - n)$
    & $L^{\alpha}_n(x)= C_n\exp{\left[-x\frac{d^2}{dx^2}
    -(\alpha+1)\frac{d}{dx}\right]}. x^n$ \\
&&    \\
$\left[x^2\frac{d^2}{dx^2}+2x\frac{d}{dx}-n(n+1)-\frac{d^2}{dx^2}\right]P_n(x)=0$
&$(D+n+1)(D-n)$ & $P_n(x)= C_n \exp{\left[-\frac{1}{2(D+n+1)}
    \frac{d^2}{dx^2}\right]}. x^n$ \\
$\left[x^2\frac{d^2}{dx^2}+(2\lambda+1) x\frac{d}{dx}-
n(2\lambda+n)-\frac{d^2}{dx^2}\right]$$\times$
&$(D+n+2\lambda)(D-n)$ & $C_n^\lambda(x) = C_n \exp{
    \left[-\frac{1}{2(D + n + 2 \lambda)}
    \frac{d^2}{dx^2}\right]}. x^n$ \\
$\times$$C^{\lambda}_n(x)=0$&& \\
&&\\
$\left[x^2 \frac{d^2}{dx^2}+ x \frac{d}{dx}- n^2- \frac{d^2}{dx^2}
    \right]T_n(x)=0$
&$(D+n)(D-n)$ & $T_n(x)=C_n\exp{\left[-\frac{1}{2(D+n)}
    \frac{d^2}{dx^2}\right]}.x^n$\\
$\left[x^2\frac{d^2}{dx^2}+3x\frac{d}{dx}-n(n+2)-\frac{d^2}{dx^2}\right]U_n(x)=0$
&$(D+n+2)(D-n)$
&$U_n(x)=C_n\exp{\left[-\frac{1}{2(D+n+2)}\frac{d^2}{dx^2}
    \right]}. x^n$\\
&&    \\
$\left[x^2\frac{d^2}{dx^2}+x\frac{d}{dx}- \nu^2 + x^2
\right]J_{\pm \nu}(x)=0$ &$(D+\nu)(D-\nu)$ &$J_{\pm \nu}(x)=C_{\pm
    \nu}\exp{\left[-\frac{1}{2(D \pm \nu)}x^2\right]}. x^{\mp\nu} $\\
&&\\
$\left[x \frac{d}{dx}+\alpha - x \frac{d^2}{dx^2} -
    \gamma\frac{d}{dx}\right] \Phi(\alpha,\gamma,x)=0$
&$(D+\alpha)$ &$\Phi(\alpha,\gamma,x)= C_{-\alpha} \exp{\left[-x
    \frac{d^2}{dx^2}-\gamma \frac{d}{dx} \right]}. x^{- \alpha}$\\
&&    \\
$\left[x(1-x)\frac{d^2}{dx^2}+(\gamma-[\alpha+\beta+1]x)\frac{d}{dx}
    -\alpha\beta\right]\times$
& $(D+\alpha)(D+\beta)$
&$F(\alpha,\beta,\gamma,x)=C_{- (\alpha,\beta)}\times$\\
&&\\
$\times F(\alpha,\beta,\gamma,x)=0$&
&$\times\exp{\left[-\frac{1}{[D+(\alpha,\beta)]}(x\frac{d^2}{dx^2}+\gamma\frac{d}{dx})
    \right]}. x^{-(\beta,\alpha)}$\\
&&\\
\hline \hline
\end{tabular}
\end{center}
The solution to the DE $\left[F(D) + P(x,d/dx)\right] y(x) = 0 $, is,\\
$y(x) = C_\lambda \left\{\sum_{m = 0}^{\infty} (-1)^m
\left[\frac{1}{F(D)}P(x,d/dx)\right]^m \right\} x^\lambda $,
provided $F(D)x^{\lambda}=0$.  \\

\end{document}